\documentstyle[12pt]{article}
\def\be{\begin{equation}}
\def\ee{\end{equation}}
\def\s{\sigma}
\def\de{\delta}
\def\om{\omega}
\def\pr{\prime}
\title{Excitations of rotational states\\
in string model of meson\\
and in three-string baryon model }
\author{G.\,S. Sharov\thanks{E-mail: german.sharov@tversu.ru}}
\date{}
\begin{document}
\maketitle
\centerline{\small
Tver State University,
Sadovyj per., 35, 170002, Tver, Russia.}
\vspace{3mm}

\begin{abstract}
For the relativistic string with massive ends, the three-string baryon model
and other string models of the baryon the small disturbances of well known
rotational motions are considered. For the string meson model the two types
of these disturbances (quasirotational motions) are obtained.
They are oscillatory motions in the form of stationary
waves in the rotational plane and in the orthogonal direction.
These solutions may be used  for describing more wide spectrum of
hadron excitations and for the quantization of this nonlinear system.
For three-string baryon model or Y configuration these quasirotational motions
are more complicated and contain the branch with growing amplitudes.
This consideration proves that for the three-string model with both massless
and massive ends the rotational motions are unstable on the classical level.
\end{abstract}

\bigskip
\noindent{\bf Introduction}
\medskip

The considered string meson model \cite{Ch,BN} includes two massive
points connected by the relativistic string that simulates the QCD confinement
mechanism. String models of the baryon were suggested in Ref.~\cite{AY}
shortly after the meson one in the following four variants
with different topology of spatial junction of three massive points (quarks)
by relativistic strings:
(a) the meson-like quark-diquark model $q$-$qq$ \cite{Ko},
(b) the linear configuration $q$-$q$-$q$ \cite{lin},
(c) the ``three-string" model or Y configuration \cite{Collins,PY},
and (d) the ``triangle" model or $\Delta$ configuration \cite{Tr,PRTr}.

The problem of choosing the most adequate string baryon model among
the four mentioned ones has not been solved yet. All these models can work in the particle physics and under certain assumptions describe the linearly
growing Regge trajectories for orbitally excited
baryon states \cite{4B} (Sect.~1).
All models have a some degree of the QCD motivation but, in particular,
different authors in the frameworks the QCD-based baryon Wilson loop operator
approach give some arguments in favour of the Y configuration \cite{Kalash} or
the ``triangle" model \cite{Corn}.

The quark-diquark configuration was more popular (in comparison with other
string baryon models) because of its simplicity and similarity with the meson
model.  Any approach suggested for the meson string model $q$-$\overline q$
may be applied to the baryon model $q$-$qq$ with minimal corrections.
The important advantage of the quark-diquark string baryon configuration
(this concerns also to the $q$-$q$-$q$ model) is connected with the equality
of the predicted Regge slopes for mesons and baryons if the string tensions
ear equal.

The linear $q$-$q$-$q$ configuration was supposed to be unstable with respect
to transformation into the quark-diquark one \cite{Ko}.
However this problem was studied quantitatively
only recently in Ref.~\cite{lin}, where we showed that the
classic rotational motions of this model are unstable indeed, but the system
doesn't transform into the $q$-$qq$ configuration.
For the three-string model or Y configuration the massless variant was better
investigated \cite{AY,Collins,PY} in comparison with $q$-$q$-$q$ or $\Delta$
ones (for other models the massless case is trivial non-baryon open or closed
string), but even for the massless three-string the dynamics remains too
complicated. The delay with study of the baryon model ``triangle" was
also connected with some difficulties in its dynamics.

To solve the stability problem for the classic rotations for all
mentioned models is very important for choosing the most adequate string
baryon configuration. This problem was studied in Ref.~\cite{stabPRD}.
This was surprising, but the rotations of the Y configuration appeared
to be unstable too. In Sect.~3 of this paper the later instability
is proved (even in the linear approximation) with the help of analysis
of the quasirotational motions of the three-string model.
This result is compared with the stability of the rotational motions
for the relativistic string with massive ends that was
obtained in Sect.~2 in the linear approximation.
The latter result gives a possibility for quantization of this
nonlinear system in the linear vicinity of the rotational solution
(this is possible only for a stable solution).

\bigskip
\noindent{\bf 1. Rotational states and Regge trajectories}
\medskip

For all mentioned string hadron models the classical solutions
describing the rotational motion (planar uniform rotations of the
system) are known and widely used for modeling the orbitally excited hadron
states on the main Regge trajectories \cite{Ko,4B,InSh,Solov}.
The rotational motion of the meson model or the baryon configuration
$q$-$qq$ is a rotation of the rectilinear string segment \cite{Ch,Ko}
(with the middle quark at the rotational center for the model $q$-$q$-$q$).
The world surfaces $X^\mu(\tau,\sigma)$ of these rotational motions
may be represented in the form \cite{4B,stabPRD}
\be
X^\mu(\tau,\sigma)=X^\mu_{rot}(\tau,\sigma)=\Omega^{-1}\big[\om\tau e_0^\mu+
\cos(\om\s+\phi_0)\cdot e^\mu(\tau)\big].
\label{rot}\ee
Here $\Omega$ is the angular velocity,
$e_0^\mu$ is the unit time-like velocity vector of c.m. in Minkowski
space with signature $+,-,-,\dots$,
$e^\mu(\tau)=e_1^\mu\cos\om\tau+e_2^\mu\sin\om\tau$ is the unit ($e^2=-1$)
space-like rotating vector directed along the string,
$\s\in[\s_1,\s_2]$, where $\s_1=0$, $\s_2=\pi$ without loss of generality
\cite{BN}. The parameter $\om$ (dimensionless frequency) is connected with the
constant speeds $v_1$, $v_2$ of the ends through the relations
$v_1=\cos\phi_0$, $v_2=-\cos(\pi\om+\phi_0)$,
$m_i\gamma^{-1}\Omega=(1-v_i^2)/v_i$, $\gamma$ is the string tension,
$m_i$ are the masses of the endpoints.

For the three-string model the rotational motion is also well known \cite{AY,PY}. The form of rotating Y configuration is three rectilinear
string segments with the world surfaces similar to Eq.~(\ref{rot})
joined in a plane at the angles 120${}^\circ$.

The rotational motion for the baryon model ``triangle"
describes an uniformly rotating closed string (curvilinear
triangle) composed of three segments of a hypocycloid.
The corresponding world surface is \cite{Tr,4B}
\be\textstyle
X^0=\tau-\frac TD\s,\qquad
X^1+iX^2=u(\s)\cdot e^{i\om\tau}.
\label{soltr}\ee
Here  $u(\s)=A_i\cos\om\s+B_i\sin\om\s$,
$\s\in[\s_i,\s_{i+1}]$ (for the whole world surface $\s_0<\s<\s_3$),
the six complex constants $A_i,B_i$ and the real constants $\s_i$,
 $D=\s_3-\s_0$,  $V_i^2$, $\tau^*-\tau=T$
are connected by the set of relations \cite{Tr,4B,InSh}.

Expressions (\ref{rot}) and (\ref{soltr}) are exact solutions of the classic equations
of motion for these systems satisfying all boundary conditions.
Both the equations of motion and the boundary conditions result from the
action, that for all the configurations may be presented in the form \cite{BN,4B}
\be
S=-\gamma\int\limits_G\sqrt{-g}\,d\tau\,d\s-
\sum_{i=1}^N m_i\int\sqrt{\dot x_i^2(\tau)}\,d\tau,
\label{S}\ee
where the number $N$ of material points with masses $m_i$ (quarks, antiquarks
or diquarks) is $N=2$ for the meson-like models $q$-$\overline q$, $q$-$qq$
and $N=3$ for others,
$g=\dot X^2X'{}^2-(\dot X,X')^2$,
$\dot X^\mu=\partial_\tau X^\mu$, $X^{\pr\mu}=\partial_\s X^\mu$,
the domain $G$ mapping into the world surface is bounded by the
inner lines $\s=\s_i(\tau)$ of the material points, $G$ is different for
different configurations \cite{4B,stabPRD},
$\dot x_i^\mu=\frac d{d\tau} X^\mu(\tau,\s_i(\tau))$, $c=1$.

The geometry of the systems Y and $\Delta$ should be taken into account:
for the triangle configuration the string is closed (but it is not smooth)
so we use the general form of the closure condition \cite {Tr}
$
X^\mu(\tau,\s_0(\tau))=X^\mu(\tau^*,\s_3(\tau^*))$.
Here $\s=\s_0(\tau)$ and $\s=\s_3(\tau)$
describe the trajectory of the same quark.
For the three-string baryon model in three parametrizations
$X_i^\mu(\tau_i,\s)$ of the three world sheets the different ``time-like"
parameters $\tau_i$ \cite{stabPRD} are connected at the junction world line
$\tau_2=\tau_2(\tau)$, $\tau_3=\tau_3(\tau)$, $\tau_1\equiv\tau$. So
the general form of the junction condition is
\be
X_1^\mu\big(\tau,0\big)=X_2^\mu\big(\tau_2(\tau),0\big)=
X_3^\mu\big(\tau_3(\tau),0\big).
\label{junc}\ee

The equations of motion and the boundary conditions at the
trajectories of massive points resulting from action (\ref{S})
for all the models take the simplest form under the orthonormality conditions
\be
\dot X^2+X'{}^2=0,\qquad(\dot X,X')=0,
\label{ort}\ee
which may be stated for all the models \cite{BN,Tr,stabPRD}.
If conditions (\ref{ort}) are satisfied
the equations of motion become linear
\be
\ddot X^\mu-X''{}^\mu=0,\label{eq}\ee
but the boundary conditions for the massive point at an end
\be
m_i\frac d{d\tau}U^\mu_i(\tau)\pm\gamma
\big[X'{}^\mu+\s_i'(\tau)\,\dot X^\mu\big]
\Big|_{\s=\s_i}=0,\quad\; U^\mu_i(\tau)=\frac{x_i^\mu(\tau)}{x_i^\mu(\tau)}=
\frac{\dot X^\mu+\s_i'X^{\pr\mu}}{|\dot X+\s_i'X'|}\bigg|_{\s=\s_i}\!\!
\label{qq}\ee
or in the middle point (for the models $q$-$q$-$q$ or $\Delta$)
\be
m_i\frac d{d\tau}U_i^\mu(\tau)
-\gamma\big[X'{}^\mu+\s_i'(\tau)\,\dot X^\mu\big]
\Big|_{\s=\s_i+0}
+\gamma\big[X'{}^\mu+\s_i'(\tau)\,\dot X^\mu\big]
\Big|_{\s=\s_i-0}\!\!=0,
\label{qqq}\ee
remain essentially nonlinear.
These massive points make the models much more realistic
but they bring additional nonlinearity and (hence) a lot of problems
with quantization of these models.

The energy $E$ and angular momentum $J$ of the states (\ref{rot}) and
(\ref{soltr}) are \cite{Tr,4B}
\be
E=E_{st}+\sum_{i=1}^N\frac{m_i}{\sqrt{1-v_i^2}}+
\Delta E,\quad
J=\frac1{2\Omega}\bigg[E_{st}
+\sum_{i=1}^N\frac{m_iv_i^2}{\sqrt{1-v_i^2}}\bigg]+S,
\label{EJ}\ee
where $E_{st}=\gamma\Omega^{-1}\arcsin v_i$ for the motions (\ref{rot}) and
$E_{st}=\gamma D(1-T^2/D^2)$ for the triangle states (\ref{soltr}).
The quark spins with projections $s_i$ (S=$\sum_{i=1}^Ns_i$)
are taken into account, in particular, as the spin-orbit correction
$\Delta E=\Delta E_{SL}=\sum\limits_i\beta(v_i)(\vec\om\vec s_i)$
to the energy of the classic motion. Here we use
$\beta(v_i)= 1-(1-v_i^2)^{1/2}$ for this correction \cite{4B,InSh}.
The expression (\ref{EJ}) for all string hadron models
describes quasilinear Regge trajectories with the similar ultrarelativistic
behavior:
$$
J\simeq\alpha'E^2-\nu E^{1/2}\sum_{i=1}^Nm_i^{3/2}
+\sum_{i=1}^Ns_i\big[1-\beta(v_i)\big],\quad v_i\to1.
$$
Here the slopes are different: $\alpha'=(2\pi\gamma)^{-1}$ for the meson-like
models, $\alpha'=\frac23(2\pi\gamma)^{-1}$ for the Y and
$\alpha'=\frac38(2\pi\gamma)^{-1}$ for the so called simple states (\ref{soltr}) of the triangle configuration \cite{PRTr,4B,InSh}.

The parent Regge trajectories for the $N$, $\Delta$ and strange
baryons may be described with using all string baryon models under following
assumptions: $\gamma=\gamma_{q-qq}=0.175$ GeV${}^2$, the effective
tension for the Y and ``triangle"  is to be different
$\gamma_Y=\frac23\gamma$, $\gamma_\Delta=\frac38\gamma$.
Under these assumptions and the effective quark masses $m_u=m_d=130$ MeV,
$m_s=270$ MeV the mentioned baryonic trajectories and also the
Regge trajectories for the light and strange mesons are well described
\cite{4B,InSh}.

\bigskip
\noindent{\bf 2. Quasirotational states for the meson string model}
\medskip

After this brief review of the rotational motions we'll consider
the quasirotational states of all these systems. They are small disturbances
of the rotational motions (\ref{rot}) and (\ref{soltr}).
They are interesting due to the following three reasons:
(a) we are to search the motions describing the hadron states, which are
usually interpreted as higher radially excited states and other states
in the potential models  \cite{InSh}, in other words, we are to describe
the daughter Regge trajectories;
(b) the quasirotational states are the basis for quantization of these
nonlinear problems in the linear vicinity of the solutions (\ref{rot}),
(\ref{soltr}) (if they are stable);
(c) the quasirotational motions are necessary for solving the important problem
of stability of rotational states (\ref{rot}), (\ref{soltr}) for all mentioned string models (that has not been solved yet).

For the meson string model the quasirotational motions of slightly curved
string with massive ends were studied in Refs.~\cite{AllenOV}. But in these
papers the authors used very narrow ansatz for the search of these disturbances
and the complicated nonlinear form of the string motion equations beyond
the conditions (\ref{ort}). Besides they neglected some important dependencies
and the boundary conditions (\ref{qq}) so these solutions in
Refs.~\cite{AllenOV} were not correct (details are in Ref.~\cite{stabPRD}).

In Ref.~\cite{stabPRD} another approach for obtaining
the quasirotational solutions was suggested. It includes the
orthonormality conditions (\ref{ort}) and, hence,
the linear equations of motion (\ref{eq})
with their general solution
\be
X^\mu(\tau,\s)=\frac1{2}\big[\Psi^\mu_+(\tau+\s)+\Psi^\mu_-(\tau-\s)\big].
\label{gen}\ee
So the problem is reduced to the system of ordinary differential equations
with shifted arguments resulting from the boundary conditions (\ref{qq}).
The unknown function may be $\Psi^\mu_+(\tau)$, $\Psi^\mu_-(\tau)$, or
unit velocity vectors of the endpoints $U^\mu_1(\tau)$ or $U^\mu_2(\tau)$
--- this is equivalent due to the relations Ref.~\cite{An,PeSh}
\be
\Psi^{\pr\mu}_\pm(\tau\pm\s_i)=m_i\gamma^{-1}\big[
\sqrt{-U_i^{\pr2}(\tau)}\,U_i^\mu(\tau)\mp(-1)^i U_i^{\pr\mu}(\tau)\big].
\label{psdet}\ee

After the substitution the slightly disturbed rotational motion (\ref{rot})
\be
U^\mu_1(\tau)=U^\mu_{1rot}(\tau)+u^\mu(\tau),\qquad |u^\mu|\ll1.
\label{U+u}\ee
($U^\mu_{1rot}$ corresponds to $X^\mu_{rot}$) into the boundary conditions
(\ref{qq}), (\ref{psdet}) we obtain in the first linear approximation
the linearized system of equations describing the evolution of small arbitrary
disturbances $u^\mu$. Solutions of this system may be expanded
in the Fourier series
$u^\mu(\tau)=\sum\limits_{n=-\infty}^{+\infty}u^\mu_n\exp(-i\om_n\tau)$ and
the resulting arbitrary quasirotational motions for the string with massive
ends were obtained in the form \cite{stabPRD}
\begin{eqnarray}
&\displaystyle
X^\mu(\tau,\s)=X^\mu_{rot}(\tau,\s)+\!\sum_{n=-\infty}^\infty
\Big\{e_3^\mu\alpha_n\cos(\om_n\s+\phi_n)\exp(-i\om_n\tau)&
\nonumber\\
&\qquad\qquad{}+\beta_n\big[e_0^\mu f_0(\s)+e_\perp^\mu(\tau) f_\perp(\s)+
ie^\mu(\tau) f_r(\s)\big]\exp(-i\tilde\om_n\tau)\Big\}.&
\label{osc}
\end{eqnarray}
Here $X^\mu_{rot}$ is the pure rotational motion (\ref{rot}),
$e_\perp^\mu(\tau)={\om}^{-1}\frac d{d\tau}e^\mu(\tau)$; $e_0$, $e_1$,
$e_2$, $e_3$ is the orthonormal tetrad, $\om=\om_1$.
Each term in Eq.~(\ref{osc}) describes the
string oscillation that looks like the stationary wave with $n$ nodes.
There are two types of these stationary waves: (a) orthogonal oscillations
along $z$ or $e_3$-axis at the frequencies proportional to the roots
$\om_n$ of equation\footnote{It is interesting that the same equation
(\ref{zfreq}) describes the spectrum
of states for the relativistic string with massive ends with linearizable
boundary conditions \cite{PeSh}.}
\be
(\om^2-Q_1Q_2)\big/\big[(Q_1+Q_2)\,\omega\big]=
\cot\pi\omega,
\label{zfreq}\ee
where $Q_i=\om_1 v_i/\sqrt{1-v_i^2}={}$const, $\om_1$ is equal to the
parameter $\om$ in Eq.~(\ref{rot});
and (b) planar oscillations (in the rotational plane $e_1,e_2$)
with the dimensionless frequencies $\tilde\om_n$ satisfying the equation
\be
\frac{(\tilde\om^2-q_1)(\tilde\om^2-q_2)
-4Q_1Q_2\tilde\om^2}
{2\tilde\om\big[Q_1(\tilde\om^2-q_2)+
Q_2(\tilde\om^2-q_1)\big]}=\cot\pi\tilde\om,
\label{pfreq}\ee
Here $q_i=Q_i^2(1+v_i^{-2})$.

The frequencies $\om_n$ and $\tilde\om_n$ from Eqs.~(\ref{zfreq}) and
(\ref{pfreq}) are real numbers
so the rotations (\ref{rot}) of the string with massive ends are stable
in the linear approximation. So one may consider the expansion (\ref{osc})
for an arbitrary quasirotational motion as the basis for further quantization
of this system in the linear vicinity of the solution (\ref{rot}).

\bigskip
\noindent{\bf 3. Quasirotational states and instability
of the three-string model}
\medskip

Let us compare this picture for the quasirotational states of the string
with massive ends with the similar motions for the string baryon models
$q$-$q$-$q$, Y and $\Delta$. These motions are more complicated but,
in particular, for the three-string model they may be investigated with
using the above mentioned method of small disturbances (\ref{U+u}).
For the Y configuration \cite{Collins,PY} the three world sheets
(swept up by three segments of the relativistic string)
are parametrized with three functions $X_i^\mu(\tau_i,\s)$
with different ``time-like" parameters
$\tau_1$, $\tau_2$, $\tau_3$ connected on the junction world line ($\s=0$)
in the following manner:
$\tau_1\equiv\tau$, $\tau_2=\tau_2(\tau)$, $\tau_3=\tau_3(\tau)$
\cite{stabPRD,sty}.

The rotational motions for the three-string are parametrized similar to
(\ref{rot})
\be
X_{i(rot)}^\mu(\tau_i,\sigma)=\Omega^{-1}\big[\om\tau_i e_0^\mu+
\sin\om\s\cdot e^\mu(\tau_i+\Delta_i)\big],
\quad\Delta_i=2\pi(i-1)/(3\omega_1).
\label{yrot}\ee
Substituting small disturbances (\ref{U+u})
$U_i^\mu(\tau_i)=U_{i(rot)}^\mu(\tau_i)+u_i^\mu(\tau_i)$
into the boundary conditions and searching solutions of this linearized system
in the form
$$
u_i^\mu(\tau_i)=\big[A_i^0e_0^\mu+A_ie^\mu(\tau_i+\Delta_i)+
v_i^{-1}A_i^0 e_\perp^\mu(\tau_i+\Delta_i)+A_i^ze_3^\mu\big]
\exp(-i\tilde\om\tau_i)$$
with $\tau_i(\tau)=\tau+\delta_i\exp(-i\tilde\om\tau)$
we obtain the linear homogeneous system with respect to amplitudes
$A_i^0$, $A_i$, $A_i^z$, $\delta_i$. It's nontrivial solutions
similar to Eq.~(\ref{osc}) describe separately quasirotational motions
in $e_3$ or $z$-direction and planar motions. In the first case
the equations for amplitudes $A_i^z$ are
\be
(\om\cot\pi\om+Q_1)(A_1^z+A_2^z+A_3^z)=0,\quad
(Q_1\cot\pi\om-\om)(A_1^z-A_i^z)=0,\quad i=2,3.
\label{zeq}\ee
Here we consider for simplicity the system with equal masses $m_i=m_1$
and $Q_i=\om_1 v_i/\sqrt{1-v_i^2}=Q_1$.
Nontrivial solutions of Eq.~(\ref{zeq})
are possible only if $\om=\om_n$ are roots of the equations
\be
\om/Q_1=\cot\pi\om,\qquad
\om/Q_1=-\tan\pi\om.
\label{zyfreq}\ee
The first equation coincides with Eq.~(\ref{zfreq}) in the case $m_2\to\infty$
or $Q_2=0$. The quasirotational states corresponding to each root of this
equation $\om=\om_n$, $n=1,2,\dots$ are
\be
X_i^\mu(\tau_i,\s)=\check X_{i(rot)}^\mu+|A_i^z|\sqrt{1+Q_1^2/\om_n^2}\,
e_3^\mu\sin\om_n\s\cdot\cos(\om_n\tau+\psi_n).
\label{zwaves}\ee
Here $\tau_i=\tau$ and for each $n$ we have two independent modes
of the oscillations (stationary waves) with the amplitudes $A_i^z$
connected by the relation $A_1^z+A_2^z+A_3^z=0$.

For the second equation (\ref{zyfreq}) corresponding the stationary waves
are described by the same relation (\ref{zwaves}) but with $\cos\om_n\s$
instead of $\sin\om_n\s$ and by the equality $A_1^z=A_2^z=A_3^z$. That
is the oscillations of three string segments are synchronous and equal,
the junction oscillates too. For all modes the roots of Eqs.~(\ref{zyfreq})
are real numbers.

The planar quasirotational motions of the three-string are more complicated.
The system for the amplitudes $A_i^0$, $A_i$, $\de_i$ similar to
Eqs.~(\ref{zeq}) consists of 8 independent equations
$$
\begin{array}{c}
i\tilde\om\Gamma\big[Q_1(C_{\tilde\om}+i)\,\de_i+C_{\tilde\om} (A_1-A_i)\big]
+\big[Q_1(1+v_1^{-2}) C_{\tilde\om}-\tilde\om\big](A_1^0-A_i^0)=0,\\
\epsilon_iv\Gamma^2(\tilde\om^2-\om^2)\,A_i+2K_1A_1^0+
(K_1-i\epsilon_iK_2)\,A_i^0 =0,\rule[3mm]{0mm}{1mm}\\
2K_1A_1+(K_1+i\epsilon_iK_2)\,A_i+\epsilon_iK_3A_i^0=0,\rule[3mm]{0mm}{1mm}\\
i\tilde\om\Gamma(A_1+A_2+A_3)+\big[Q_1(1+v_1^{-2})+\tilde\om
C_{\tilde\om}\big](A_1^0+A_2^0+A_3^0)=0,
\rule[3mm]{0mm}{1mm}\\
K_4\big[2\tilde\om A_1^0-(\tilde\om-i\sqrt3\,\om_1)\,A_2^0-
(\tilde\om+i\sqrt3\,\om_1)\,A_3^0\big]=0,\rule[3mm]{0mm}{1mm}\\
\big[K_4(\tilde\om^2-\om_1^2)+2iK_1(\tilde\om^2+\om_1^2)\big](A_2^0-A_3^0)=0.
\rule[3mm]{0mm}{1mm}
\end{array}
$$
Here in the first 3 equations $i=2,3$,
$C_{\tilde\om}=\cot\pi\tilde\om$, $\epsilon_i=(-1)^i\sqrt3$,
$q_1=Q_1^2(1+v_1^{-2})$, $K_1=q_1-\tilde\om^2+2Q_1\tilde\om C_{\tilde\om}$,
$K_2=\Gamma v_1\big[(\tilde\om^2+\om_1^2)\,C_{\tilde\om}+2Q_1\tilde\om\big]$,
$K_3v_1\Gamma^2(\tilde\om^2-\om_1^2)=K_2^2-K_1^2$,
$K_4=(q_1-\tilde\om^2)\,C_{\tilde\om}-2Q_1\tilde\om$.

This system has 3 branches of
nontrivial solutions. The first two branches are described by the dimensionless
frequencies $\tilde\om=\tilde\om_n$ satisfying the equations
\be
\frac{\tilde\om^2-q_1}{2Q_1\tilde\om}=\cot\pi\tilde\om,\qquad
\frac{\tilde\om^2-q_1}{2Q_1\tilde\om}=-\tan\pi\tilde\om.
\label{pyfreq}\ee
The first Eq.~(\ref{pyfreq}) looks like from Eq.~(\ref{pfreq}) if $Q_2=0$ or
$m_2\to\infty$. These oscillations are planar symmetric stationary waves
with the amplitudes $A_1=A_2=A_3$, $A_1^0=A_2^0=A_3^0$ and with real
frequencies.

However the frequencies of the third branch are roots of the equation
\be
2\frac{Q_1\tilde\om(\om_1^2-\tilde\om^2)-i(\tilde\om^2-q_1)
(\tilde\om^2+\om_1^2)}{(\tilde\om^2-q_1)
(\tilde\om^2-\om_1^2)-4iQ_1\tilde\om(\tilde\om^2+\om_1^2)}=
\cot\pi\tilde\om,
\label{compfr}\ee
are obligatory complex numbers (except for the roots $\tilde\om=\pm\om_1$).
For example, the first roots of Eq.~(\ref{compfr}) for the motion (\ref{yrot})
with $m_i=1$, $\gamma=1$, $R_i=v_i/\Omega=0.3$ are $\tilde\om_1\simeq0.276+0.0297i$, $\tilde\om_2\simeq1.05+0.157i$,
$\tilde\om_3\simeq2.027+0.17i$, $\tilde\om_4\simeq3.0183+0.173i$.
Imaginary parts of the roots of Eq.~(\ref{compfr}) are always positive
so the disturbances of this class (branch) exponentially grow in time
in accordance with the factor
$$
\exp(-i\tilde\om_n\tau)=\exp(-i\Re\tilde\om_n\tau)\exp(\Im\tilde\om_n\tau).
$$
Arbitrary quasirotational motion (a disturbance of the rotational motion
(\ref{yrot})) may be expanded in the Fourier series with harmonics of all
classes described above. So only for the disturbances with the special
symmetry (when all amplitudes of the modes (\ref{compfr}) equal zero)
these disturbances do not grow exponentially. So we proved that the
rotational motions (\ref{yrot}) for the three-string configuration
are unstable even in the linear approximation.

The evolution of this instability was calculated numerically \cite{stabPRD,sty}
with using the developed approach of solving the initial-boundary value problem
for this model. The numerical experiments \cite{stabPRD,sty} show that
the picture of motion is qualitatively identical for any
small asymmetric disturbance.
Starting from some point in time the junction begins to move.
During this complicated motion the distance between the junction and
the rotational center increases and the lengths of the string segments vary
quasiperiodically unless one of the material points inevitably merges with the junction.

Note that for the massless ($m_i=0$) three-string model
\cite{Collins,PY} the rotational motions
are also unstable. The equation (\ref{compfr}) in this case has the form
$$2i(\tilde\om^2+\om_1^2)\big/(\om_1^2-\tilde\om^2)=\cot\pi\tilde\om,\qquad
m_i=0$$
and its roots are complex numbers.

The stability problem for the string baryon models $q$-$q$-$q$ and $\Delta$ was studied  \cite{lin,stabPRD} with solving the initial-boundary value problem
for these system. The results were different: the classic rotational motions
(\ref{soltr}) for the triangle model appeared to be stable and for the linear
$q$-$q$-$q$ configuration they are unstable.

\bigskip
\noindent{\bf Conclusion}
\medskip

The stability problem for classic rotations is solved
for all string meson and baryon models.
Rotational motions (\ref{rot}) for the relativistic string with massive ends
are classically {\it stable} in the linear approximation.
Any small disturbances of these motions
may be represented in the form of Fourier series (\ref{osc}).
 Rotational motions (\ref{rot}) and (\ref{yrot}) for the string baryon models
$q$-$q$-$q$ and Y are {\it unstable} on the classic level (any asymmetric
disturbances of them grow exponentially), but the similar motions
(\ref{soltr}) of the triangle configuration are {\it stable}.
Instability of the state (\ref{yrot}) for the three-string baryon model
is connected with the presence of the modes with growing amplitudes
or complex frequencies (\ref{compfr}) in the spectrum of quasirotational
excitations.

 Quasirotational motions of the string with massive ends in the
form  (\ref{osc}) may be used as the basis of quantization
in the linear vicinity of the stable solution (\ref{rot}). Expression
(\ref{osc}) satisfies the constraint (\ref{ort}) so we have no an analog
of the Virasoro conditions in this case.
 Progress in quantization of the considered nonlinear meson and baryon
string models is necessary for describing not only orbital but also radial
and other excitations of hadrons known from potential models
\cite{InSh}.
 Quantization in the linear vicinity is possible only for stable
solutions, so we have some problems here for the linear
$q$-$q$-$q$ and three-string models. However
this doesn't mean that these models are finally closed for further
applications.

\noindent{\bf Acknowledgement.}
The work is supported by the Russian Foundation of Basic Research
(grant 00-02-17359).

\newcommand{\etal}{{\em et al.}}

\end{document}